\documentclass[manuscript,revised,noblind]{geophysics}
\renewcommand{\today}{April 25, 2026}

\usepackage{array}
\usepackage{multirow}
\usepackage{tabularx}
\usepackage{amsmath,amssymb,amsfonts}

\begin{document}
\title{Pretrain-to-alignment learning paradigm to improve geophysical AI applicability under scarce field labels and synthetic-to-field gaps: A case study of relative geologic time estimation in global shelf-edge clinothems}

\address{
\footnotemark[1] School of Earth and Space Sciences, State Key Laboratory of Precision Geodesy, Mengcheng National Geophysical Observatory, University of Science and Technology of China, Hefei 230026, China \\
\footnotemark[2]School of Artificial Intelligence, State Key Laboratory of Digital Intelligent Technology for Unmanned Coal Mining, Anhui University of Science and Technology, Hefei 231131, China}
\author{Hui Gao\footnotemark[1], Xinming Wu\footnotemark[1], Jiarun Yang\footnotemark[2], Zhixiang Guo\footnotemark[1], and Yimin Dou\footnotemark[1]}

\footer{Example}
\lefthead{Gao et al.}
\righthead{Pretrain-to-alignment learning paradigm}
\renewcommand{\thefootnote}{\fnsymbol{footnote}} 
\renewcommand{\figdir}{figures} 

\begin{abstract}
Artificial intelligence (AI) has been increasingly applied to various geophysical scenarios, yet its practical deployment remains limited by scarce field labels, pronounced synthetic-to-field domain gaps, and insufficient physical consistency under complex and variable field conditions. To address these challenges, we propose a pretrain-to-alignment learning paradigm that systematically integrates self-supervised pretraining, synthetic supervision, prior-driven refinement, and domain-adaptation fine-tuning into a unified progressive learning workflow. In this paradigm, geophysical AI models are developed through sequential stages that progressively build field-relevant representations, task-specific mapping capability, field consistency, and target-specific adaptability. We validate this paradigm using cross-survey relative geologic time (RGT) estimation in global shelf-edge clinothems as a representative case study. Results from 3,000 field datasets spanning multiple sedimentary basins demonstrate that the proposed paradigm achieves accurate, robust, and well-generalized performance across diverse field surveys, while significantly improving fine-scale stratigraphic and structural details. More broadly, this study provides a practical methodological reference for a broader range of geophysical AI tasks, such as interpretation, regression, and inversion problems.
\end{abstract}

\section{Introduction}
Artificial intelligence (AI) approaches have been widely applied across numerous geophysical tasks, including feature detection and delineation~\citep{xiong2018seismic,wu2019faultseg3d,shi2019saltseg,pham2019automatic,gao2025foundation}, quantitative regression and property estimation~\citep{das2019convolutional,geng2020deep,das2020petrophysical,kazei2021mapping,han2025self}, inverse imaging and reconstruction~\citep{araya2018deep,yang2019deep,wu2019parametric,li2021deep,adler2021deep}, and event detection and monitoring~\citep{mousavi2016seismic,zhu2019phasenet,mousavi2020earthquake,anikiev2023machine,si2024all}. Despite these advances, robust cross-survey generalization and reliable field deployment remain major challenges. First, high-quality labeled field datasets are scarce and expensive to acquire~\citep{an2023current,lin2024machine,liu2024foundation,xu2025machine}, which limits the feasibility of large-scale supervised learning in real applications. Second, although synthetic datasets can provide abundant and reasonable labels, they often differ substantially from field data in data distribution, structural complexity, and realism, leading to an unavoidable and pronounced synthetic-to-field domain gap~\citep{zhou2021learning,farris2023learning,zhou2025fault}. Third, purely data-driven AI methods often suffer from overfitting to statistical patterns, sensitivity to domain variability, insufficient physical consistency when applied to complex field data~\citep{zhou2019seismic,zhao2024artificial,xu20253d}. Fourth, the trained model often exhibits limited cross-survey generalization and requires case-specific retraining or additional adaptation before reliable practical deployment~\citep{zhao2024artificial,yang2025seismic,quesada2025large}. Collectively, these issues have become primary bottlenecks to the practical deployment of geophysical AI across various field surveys and geological settings.

To improve the practical applicability of geophysical AI models, existing studies have advanced several major directions, including few-shot or transfer learning with limited labeled field data~\citep{cunha2020seismic,zhao2023few,liu2025deep}, synthetic-data-driven supervision to mitigate label scarcity~\citep{wu2020building,zhao2023fault2seisgan,gao2023clinoformnet}, large-scale pretraining on unlabeled data for improving feature representations~\citep{liu2024foundation,sheng2025seismic,chu2025pretraining,dou2025geological}, domain knowledge or physical constraints-driven model training~\citep{wu2023sensing,zuo2024novel,cao2024embedded,xu2025geological}, and fine-tuning or domain adaptation for target-specific transfer~\citep{zhou2021learning,gao2025geologically,yang2025seismic}. Although these strategies contribute to enhancing the model's capability and applicability, they are typically developed in parallel and implemented separately, therefore remain insufficient to constitute a comprehensive solution for practical deployment. Few-shot or transfer learning using labeled field data is inherently constrained by the subjectivity and expert-dependence of manual annotations, the limited scale and diversity of labeled datasets, and overfitting to specific patterns, leading to weak generalization across surveys. Synthetic-data-driven supervision contributes to mitigate the absence of labels, but models trained solely in this way often remain sensitive to the synthetic-to-field domain gap. Large-scale pretraining can improve field-relevant feature representations, yet such improvements do not necessarily translate into stable and reliable downstream performance without sufficient task-specific supervision. Similarly, knowledge-guided training or target-specific transfer strategies can enhance prediction plausibility and adaptability, but they commonly require a sufficiently strong initialization model and are often applied as independent refinements rather than being systematically incorporated into a complete training-to-deployment framework. Consequently, current geophysical AI research still lacks a comprehensive and transferable learning paradigm that systematically integrates representation learning, synthetic supervision, knowledge-driven refinement, and deployment-stage adaptation.

To address these limitations, we propose a pretrain-to-alignment learning paradigm for improving the applicability of geophysical AI under scarce field labels and pronounced synthetic-to-field domain gaps. Instead of treating model development as a single-stage optimization problem, the proposed paradigm organizes it as a unified progressive learning process that systematically integrates self-supervised pretraining, synthetic supervision, geological prior-driven refinement and lightweight domain-adaptation fine-tuning (Figure~\ref{fig:fig1_1}). In this framework, the geophysical AI model initially acquires field-relevant representations from large-scale unlabeled field datasets through self-supervised pretraining, and then learns task-specific mappings from massive-scale labeled synthetic datasets via synthetic supervision. Subsequently, the trained model is further refined on large-scale unlabeled field datasets using label-independent prior constraints or regularization terms, thereby progressively aligning its predictions with field conditions and reducing the synthetic-to-field gap. Finally, lightweight domain-adaptation fine-tuning is performed on an individual target sample to account for target-specific variability and enhance reliability and flexibility in practical applications. In this way, the proposed paradigm establishes a complete and transferable learning framework for progressively improving model applicability and bridging the synthetic-to-field gap from representation learning to deployment-stage adaptation.

To demonstrate the effectiveness of the proposed paradigm, we use relative geologic time (RGT, ~\cite{stark2003unwrapping,stark2004relative}) estimation in global shelf-edge clinothems as a representative case study. Clinothems are basinward-accreting depositional prisms formed at sedimentary basin margin and constitute fundamental elements of the stratigraphic archives ~\citep{steel2002sequence,hampson2003geomorphological,patruno2018clinoforms}. Their geometries and stacking patterns record the migration of depositional environments and are primarily forced by variations in sediment supply and relative sea level, which control the progradation, aggradation, and retrogradation of shelf edge in both siliciclastic and carbonate systems ~\citep{pirmez1998clinoform,bullimore2005clinoform,posamentier2007seismic,anell2017quantifiable,patruno2018clinoforms}. Therefore, accurate RGT estimation in such clinothem-dominated systems is critical for stratigraphic framework construction, stacking pattern interpretation, geomorphological and geological structural analysis, unconformities and erosional surfaces identifications, sand-bodies characterization, reservoir property modeling, and sedimentary evolutionary history reconstruction ~\citep{sydow2013stacked,holgate2014constraining,allen2013basin}. Meanwhile, it remains a highly challenging continuous regression task that requires AI model to simultaneously capture large-scale stratigraphic architecture, fine-scale reflector geometry, stratal terminations, and unconformity-related complexities. In particular, successful prediction depends on robust representation learning, reliable task-specific mapping, effective geological alignment across different depositional settings, and robust adaptation to target-specific field variability. These requirements make estimating RGT of global shelf-edge clinothems a particularly suitable and demanding case study for evaluating the proposed pretrain-to-alignment paradigm.

\section{Method and Data}
\subsection{Pretrain-to-Alignment Learning Paradigm} \label{sec2_1}
\plot{fig1_1}{width=\textwidth}
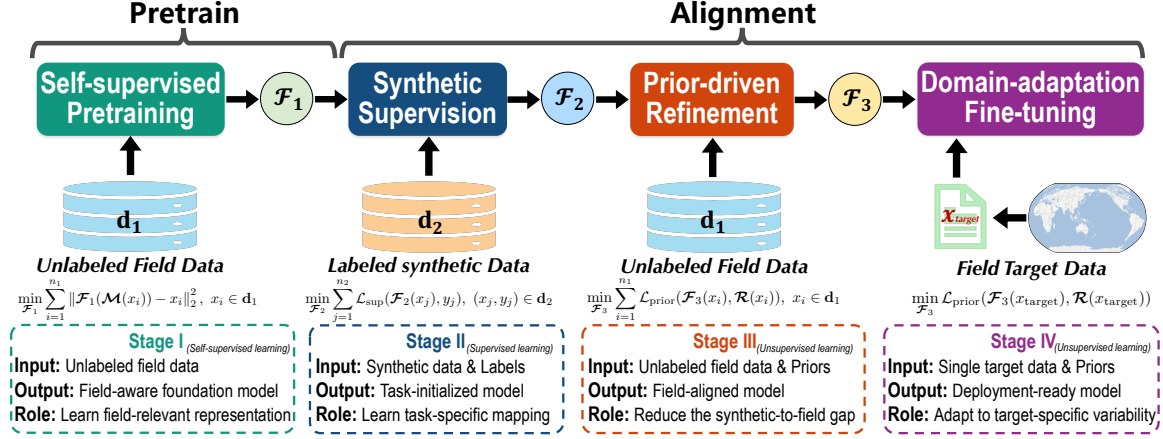
{Conceptual overview of the proposed pretrain-to-alignment paradigm. The paradigm consists of four progressive learning stages, namely self-supervised pretraining, synthetic supervision,  prior-driven refinement, and domain-adaptation fine-tuning. Through the progressive integration of these stages using unlabeled field data ($d_1$), labeled synthetic data  ($d_2$), and individual target data ($x_{target}$), the AI model gradually evolves from representation learning toward alignment with field conditions and target-specific variability, thereby reducing the synthetic-to-field gap and improving deployment reliability.}

In this study, we propose a pretrain-to-alignment learning paradigm for geophysical AI under scarce field labels and pronounced synthetic-to-field domain gaps. As illustrated in Figure~\ref{fig:fig1_1}, the proposed paradigm establishes a comprehensive and transferable learning route that systematically integrates self-supervised pretraining, synthetic supervision, prior-driven refinement, and domain-adaptation fine-tuning, thereby progressively and robustly the perception capability and practical applicability. 
Conceptually, the proposed paradigm can be divided into an initial large-scale pretraining stage and three subsequent alignment stages for task-initialization, field alignment, and deployment adaptation. In the first stage, self-supervised pretraining is performed on large-scale unlabeled field data, enabling the AI model to learn field-relevant structural patterns, statistical characteristics, and multi-scale features without requiring manual annotations. This stage provides a robust field-aware representational foundation and an effective model initialization for subsequent downstream tasks. In the second stage, synthetic supervision is introduced on massive labeled synthetic data to mitigate the scarcity of field annotations and provide realistic and controllable labels for downstream tasks, thereby enabling the AI model to establish an initial task-specific mapping under explicit supervision. This stage converts the field-aware representational capability learned in the first stage into a robust task-initialized model for the downstream task and establishes the basis for subsequent synthetic-to-field alignment.

Since the task-initialized model trained solely under synthetic supervision remains insufficiently adapted to field data, we then perform prior-driven refinement using large-scale unlabeled field data to align the learned task-specific mapping with field conditions and characteristics. By introducing label-independent prior constraints or regularization terms derived from field data, the task-initialized model is gradually refined under field data distributions, thereby reducing the synthetic-to-field gap, improving physical consistency, and transforming it into a field-aligned model. In the final stage, we apply lightweight domain-adaptation fine-tuning to an individual target sample to further adapt the field-aligned model to target-specific variability and improve reliability during practical deployment. Unlike the previous stages, which rely on large-scale training datasets, this stage emphasizes deployment-oriented adaptation at the sample level. Through directly optimizing the target-sample prediction using objectives driven from label-independent prior constraints or regularization terms, the final model can accurately capture target-specific characteristics and achieve more consistent and stable deployment performance.

Overall, the proposed paradigm treats the geophysical AI model development as a progressive capability-building process rather than a single-stage optimization problem. Within this framework, the AI model does not attempt to completely solve the downstream task at once, but instead gradually acquires field-aware representations, task-specific prediction ability, field-aligned refinement, and target-specific adaptability. In this way, it progressively addresses several key bottlenecks in geophysical AI, including scarce field labels, synthetic-to-field domain gap, weak physical consistency, limited cross-survey generalization, and unstable deployment performance. The value of this paradigm lies not simply in combining multiple learning strategies, but in organizing them into a coherent and complementary process through which different model capabilities are progressively established, refined, and adapted under field conditions. Consequently, the proposed paradigm provides a general methodological framework that can be readily extended a broader range of geophysical AI tasks, including interpretation, regression, and inversion problems, while improving the model applicability under complex and variable field deployment conditions.

\subsection{Dataset for Cross-survey RGT Estimation in Global Shelf-edge Clinothems} \label{sec2_2}
To instantiate and evaluate the proposed learning paradigm, we use a hybrid benchmark dataset curated by \cite{gao2026dataset} for global shelf-edge clinothems. This dataset comprises 3,000 unlabeled field samples and 4,000 labeled synthetic samples with corresponding RGT labels. The unlabeled field dataset was constructed through a standardized field data curation workflow, including publicly available data collection, manual selection, data cropping, resampling, and standardization. These field samples were collected from multiple representative sedimentary basins worldwide, including the North Sea, Taranaki, Northern Carnarvon, and Beaufort–Mackenzie basins, and cover a broad range of depositional settings, stratigraphic architectures, and clinothem geometries. Therefore, the field dataset preserves the realism, diversity, and structural complexity of shelf-edge clinothem-dominated systems, and provides a realistic and representative data basis for self-supervised pretraining, prior-driven refinement, and cross-survey evaluation under complex geological conditions.

The labeled synthetic dataset was simulated through a three-step geological and geophysical forward modeling workflow, including stratigraphic forward modeling, geological structural deformation, and geophysical forward modeling. This workflow generates massive-scale geologically plausible synthetic seismic samples and corresponding dense and structurally consistent RGT labels, which provide explicit supervision for learning a stable seismic-to-RGT mapping. Compared with the realistic but unlabeled field dataset, the synthetic dataset provides controllable, label-complete, and geologically consistent training samples for supervised training and quantitative assessment. Collectively, these two complementary datasets establish an effective benchmark foundation and a representative application scenario for developing, validating, and practical assessing the proposed learning paradigm in complex global shelf-edge clinothem-dominated systems. 
 

\section{Results}
\plot{fig1_2}{width=\textwidth}
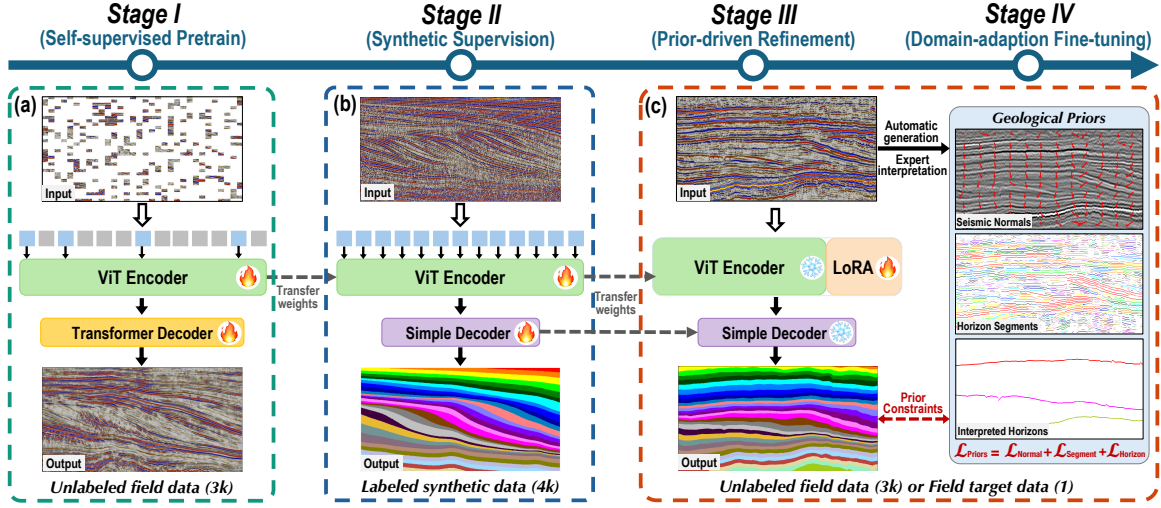
{Geological prior-driven pretrain-to-alignment AI framework for RGT estimation in global shelf-edge clinothems. (a) In the first stage, we perform continued self-supervised pretraining of a seismic foundation model~\citep{sheng2025seismic} on 3,000 unlabeled field data to learn field-aware and clinothem-sensitive seismic representations. (b) In the second stage, the pretrained ViT encoder module is transferred to an RGT prediction network, where a simple convolutional decoder with four convolutional blocks is introduced and trained on 4,000 labeled synthetic data to establish an initial seismic-to-RGT mapping. (c) In the subsequent alignment stages, the pretrained backbone is frozen and lightweight LoRA ~\citep{hu2021lora} modules are inserted into the encoder for parameter-efficient refinement and adaptation. Specifically, the 3,000 unlabeled field data are reintroduced for model fine-tuning by formulating multiple geological priors as label-independent loss functions to directly regularize the predicted RGT field, thereby enhancing geological consistency and cross-survey generalization. Finally, these geological prior losses are further analogously treated as optimization objectives for sample-wised adaptation on an individual field sample, thus enabling domain-adaptation fine-tuning and improving deployment reliability.}

In this section, we instantiate the proposed pretrain-to-alignment learning paradigm through a representative case study of RGT estimation in global shelf-edge clinothems. This regression task provides a representative and challenging test scenario, which requires AI models to simultaneously capture large-scale geological structures, fine-scale reflector geometry, geological consistency, and robustness to variability across sedimentary basins. Following the progressive learning framework described in Sec.\ref{sec2_1}, we first preform self-supervised pretraining on the unlabeled field data, then establish an initial seismic-to-RGT mapping through synthetic supervision on the labeled synthetic data, and subsequently improve field alignment and sample-specific adaptability through prior-driven refinement and domain-adaptation fine-tuning (Figure~\ref{fig:fig1_2}). Through these progressive capability-building stages, the final AI model achieve accurate, robust, and well-generalized RGT estimation across multiple geological surveys, while retaining the flexibility for efficient adaptation to individual field sample.

\subsection{Foundation Model with Continued Self-supervised Pretraining}

\plot{fig1_3}{width=\textwidth}
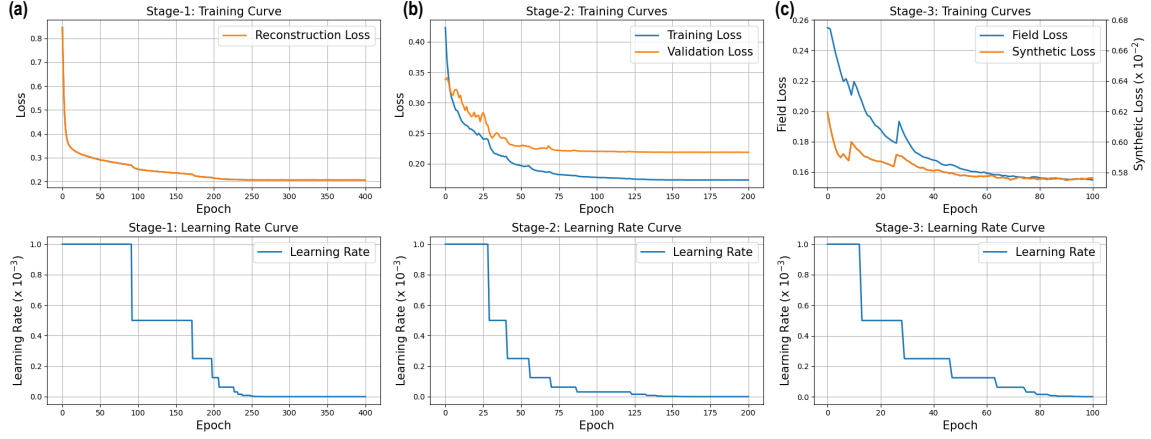
{The loss and learning rate curves across different training stages. (a) In the first self-supervised training stage, the reconstruction loss converges to 0.206 after 400 epochs, while the learning rate gradually decays from 0.001 to below $10^{-8}$. (b) In the second synthetic supervised training stage, the training and validation losses stably converge to 0.17 and 0.21 after 200 epochs, respectively. (c) In the third unsupervised training stage, the field loss gradually decays to 0.15, while the synthetic loss remains at a consistently low level.}

In the self-supervised pretraining stage (see Figure~\ref{fig:fig1_2}a), we initialize the network's parameters using a publicly available seismic foundation model ~\citep{sheng2025seismic}, which was pretrained on nearly two million diverse seismic images. However, clinothem-dominated samples are substantially underrepresented in the original pretrained dataset, leading to feature representations biased toward more common seismic patterns, such as parallel, subparallel, or chaotic facies. Therefore, the original pretrained model is less sensitive to clinothem-specific characteristics that are critical for our task, including seismic reflector continuity, stratigraphic stacking patterns, and stratal terminations such as toplap, onlap, downlap, and truncation.

To mitigate this limitation, we perform task-oriented self-supervised re-pretraining using the masked autoencoder (MAE) framework \citep{he2022masked} on the unlabeled field dataset \citep{gao2026dataset}, which contains diverse clinothem architectures. We train the network using a masked mean squared error (MSE) reconstruction loss under a gradually decaying learning rate schedule. As shown in Figure~\ref{fig:fig1_3}a, the reconstruction loss gradually converges to 0.206 after 400 epochs, while the learning rate progressively decreases from 0.001 to below $10^{-8}$. By applying a high-ratio random masking strategy to the input seismic image and then reconstructing the masked regions, the network is encouraged to learn and capture both global and local structural characteristics of clinothems, thereby enhancing its perception of the clinothem geometries, stratigraphic stacking patterns, and stratal terminations. 

To evaluate the effectiveness of the re-pretraining process, we feed the same masked seismic input (Figure~\ref{fig:fig1_4}a) into both the original model and the re-pretrained model for seismic data reconstruction. Compared with the reconstructed result generated from the original model  (see Figure~\ref{fig4}b) , the re-pretrained model exhibits clear improvements in seismic reflector continuity, stratal terminations, stratal pinch-out, and unconformity regions delineations, as indicated by the yellow arrows in Figure~\ref{fig4}c. These improvements indicate that the task-oriented re-pretraining strategy significantly enhances the model's ability to characterize key structural features of clinothems, thereby providing a more suitable and structurally sensitive representation foundation for the subsequent RGT estimation task.


\plot{fig1_4}{width=0.4\textwidth}
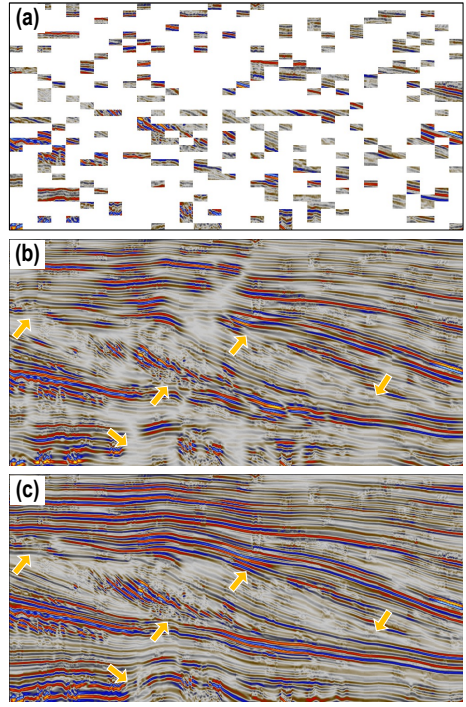
{Comparison of MAE reconstruction results using the same masked input (a). Relative to the reconstruction (b) obtained from original seismic foundation model \citep{sheng2025seismic}, the task-oriented re-pretrained model produces clearer seismic reconstruction (c), with improved reflector continuity and more distinct stratal terminations, as indicated by yellow arrows.}

\subsection{Synthetic Supervision Training}
In the subsequent supervised training stage (see Figure~\ref{fig:fig1_2}b), we retain the Vision Transformer encoder and its parameters, and introduce a simple task-specific convolutional decoder for RGT estimation on labeled synthetic seismic images \citep{gao2026dataset}. Specifically, the decoder reshapes the encoded patch tokens into 2D feature maps, progressively upsamples them through four convolutional decoding blocks, and incorporates a shallow image-level skip feature in the final decoding stage to recover fine local details. Instead of directly predicting RGT values, we formulate the output as a vertical RGT derivative field and clamp the predictions to suppress negative values and extreme outliers, which correspond to implausible tectonic inversion in the absent of reverse faults or overturned strata, and unstable artifacts, respectively. Then the constrained derivative field is vertically integrated to obtain the final RGT prediction, thereby enforcing stratigraphic sequence monotonicity and geological plausibility. Finally, the network is trained end-to-end by minimizing a composite regression loss that combines the mean squared error (MSE) and the multi-scale structural similarity (MS-SSIM) \citep{wang2003multiscale,wang2004image} between the RGT predictions and corresponding synthetic labels, enabling the network to learn a stable and geologically consistent mapping from seismic images to RGT values. During this supervised training stage, we randomly split the labeled synthetic dataset into 3,200 training samples and 800 validation samples. After 200 epochs of model training, the training and validation losses converge to 0.17 and 0.21, while the learning rate decays from 0.001 to below $10^{-6}$, as shown in Figure~\ref{fig:fig1_3}b.

We further assess the trained network by directly applying it to field seismic data shown in the first column of Figure~\ref{fig:fig1_7}. Although the model successfully captures the coarse global structural trends of the seismic image, the predicted RGT results (second column of Figure~\ref{fig:fig1_7}) exhibit limited resolution and inconsistent stratigraphic details when applied to complex geological structures. These limitations are especially evident in the key clinothems stratigraphic characteristics, including unconformity surfaces, local stratal terminations, seismic reflector continuity, and stacking patterns. These results reveal an inherent limitation of relying solely on synthetic supervision, where the unavoidable domain gap between synthetic and field seismic datasets significantly restricts the model’s fitting capability and generalizability under complex and variable field geological conditions. This issue is particularly pronounced for sensitive continuous regression tasks such as RGT estimation, where training exclusively on synthetic data is insufficient to fully capture and perceive the structural and stratigraphic complexity inherent in field datasets.

\subsection{Geological Prior-driven Refinement}
To solve the limitations in purely data-driven synthetic supervision and further improve model performance on complex field data, we introduce geological prior-driven refinement using unlabeled field datasets(see Figure~\ref{fig:fig1_2}c). In this stage, three complementary geological prior information are formulated as label-independent loss functions to directly regularize the RGT predictions, thereby achieving unsupervised refinement guided by geologically informed constraints without requiring explicit labels. We first introduce a structural orientation-alignment loss ($\mathcal{L}_{\text{Normal}}$) derived from seismic normal vectors information. Structural orientations are estimated from seismic data using the structure tensor \citep{van1995estimators,fehmers2003fast,hale2009structure}, yielding seismic normal vectors that represent local stratigraphic variation directions. We then enforce pixel-wise structural consistency between seismic normal vectors and the gradient of predicted RGT field using cosine similarity, thus aligning RGT variations with the dominant geological structural directions, as follows:\\
\begin{align}
\mathcal{L}_{\text{Normal}} = 1 - \operatorname{cosine\_similarity}( \nabla \tau, \mathbf{u} ), 
\end{align}
where $\nabla \tau$ denotes the gradient of the predicted RGT field and $\mathbf{u}$ represents the seismic normal vectors.

We further introduce a local stratigraphic isochronous loss ($\mathcal{L}_{\text{Segment}}$) based on multiple automatically grown local horizon segments. For each seismic image, we first compute several structural attributes including seismic normal vectors, linearity, and skeletonized data extracted from waveform peaks or troughs. These skeletonized points are treated as candidate seeds for local horizon growth, while the linearity attribute serves as an indicator of reflector continuity and data reliability. Starting from seed points with high linearity, a region-growing algorithm is applied to construct local horizon segments by iteratively selecting neighboring seeds that satisfy multiple conditions, including amplitude similarity, multi-scale waveforms cross-correlation consistency , and seismic normal vector alignment. Besides, we also introduce a mutual-neighbor validation strategy to ensure bidirectional optimal matching during horizon growth, where only mutually consistent seed points are allowed to merge into the same horizon segment. After applying iterative local horizon growth algorithm, we ultimately obtain multiple local horizon segments (Figure~\ref{fig:fig1_5}a) that highly consistent with seismic reflectors in the seismic image (Figure~\ref{fig:fig1_5}b). Finally, these automatically extracted horizon segments are served as local isosurface templates to impose local isochronous constraints on the predicted RGT field (Figure~\ref{fig:fig1_5}c) by enforcing RGT values within each segment to remain locally isochronous, as follows:\\
\begin{align}
\mathcal{L}_{\text{Segment}} = \frac{1}{M} \sum_{k=1}^{M} \sum_{i \in \Omega_k} \frac{ \left| \tau_{i} - \bar{\tau}_k \right| }{N_k},
\end{align}
where $\Omega_k$ denotes the $k$-th extracted horizon segment, $\tau_i$ is the predicted RGT value at pixel $i$, $\bar{\tau}_k$ is the mean RGT value within segment $\Omega_k$, and $N_k$ is the number of pixel within segment. 

Finally, we introduce a global horizon isochronous loss ($\mathcal{L}_{\text{Horizon}}$) to incorporate a limited number of complete horizons obtained from expert interpretation or automatic tracking. These sparse global horizons provide large-scale and spatially continuous stratigraphic guidance within seismic profile. Similar to the segment-based isochronous constraint, predicted RGT values along each complete expert-interpreted horizon are encouraged to remain isochronous, thereby enforcing global stratigraphic and structural consistency in the RGT field, as follows:\\
\begin{align}
\mathcal{L}_{\text{Horizon}} = \frac{1}{M} \sum_{j=1}^{M} \sum_{i \in \Gamma_j} \frac{ \left| \tau_{i} - \bar{\tau}_j \right| }{N_j},
\end{align}
where $\Gamma_j$ denotes the $j$-th interpreted complete horizon, $\tau_i$ represents the predicted RGT value at pixel $i$, and $\bar{\tau}_j$ is the mean RGT value along interpreted horizon $\Gamma_j$.

\plot{fig1_5}{width=\textwidth}
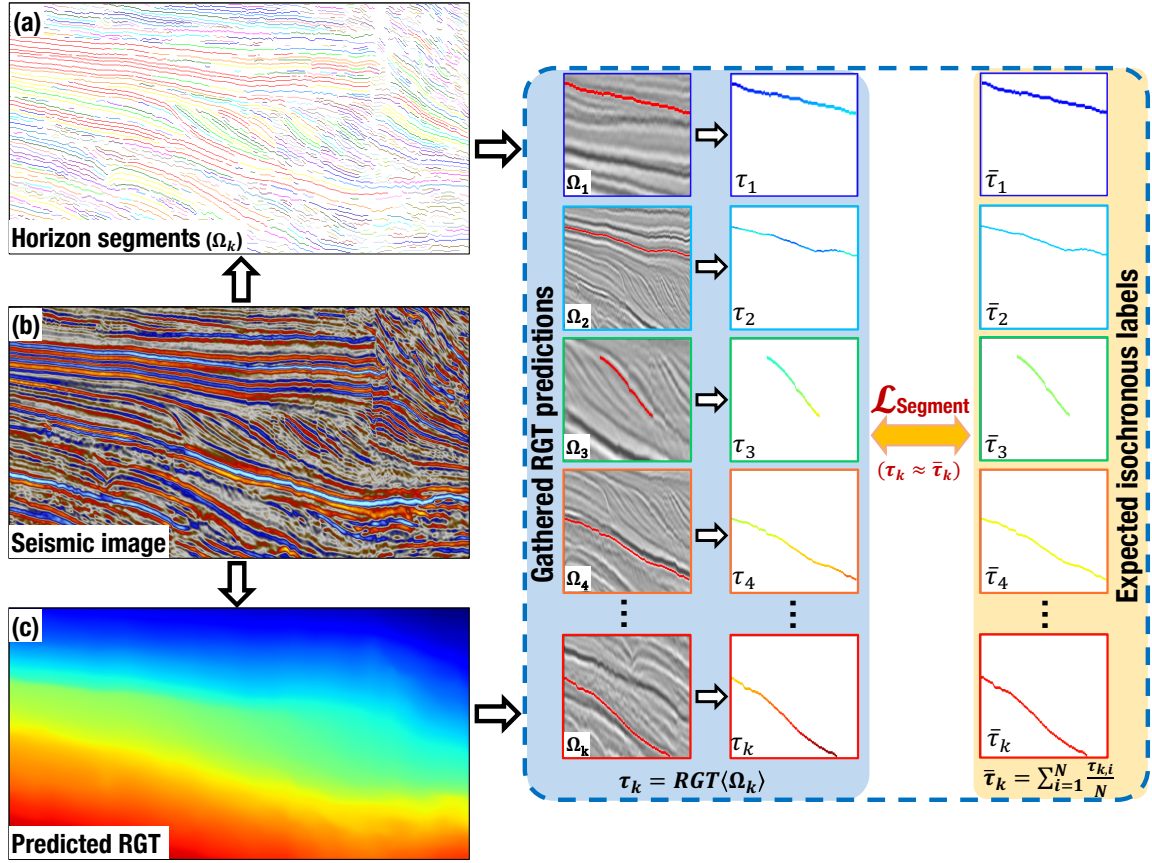
{Workflow for calculating the local stratigraphic isochronous loss ($\mathcal{L}_{\text{Isochron}}$). Specifically, we first automatically extract local horizon segments (a) from each seismic image (b) using a region-growing algorithm, where pixels with the same index belong to the same local isosurface. For each extracted horizon segment, we gather the corresponding RGT values from the predicted RGT field (c), and compute their mean values to construct the expected isochronous label. Finally, we enforce the local stratigraphic isochronism by minimizing mean absolute error (MAE) between the gathered RGT predictions with expected isochronous labels within each segment.}

In this study, we define a hybrid prior loss for geological prior-driven refinement by the following equation:\\
\begin{align}
\mathcal{L}_{\text{Priors}} = \lambda_1 \cdot \mathcal{L}_{\text{Normal}} + \lambda_2 \cdot \mathcal{L}_{\text{Segment}} + \lambda_3 \cdot \mathcal{L}_{\text{Horizon}},
\end{align}
where $\lambda_1$, $\lambda_2$, and $\lambda_3$ donate the weights of the three geological prior losses. Considering the complete expert-interpreted horizons are rarely available in field datasets, we employ only structural orientation-consistency and local segment isochronous losses ($\lambda_1=1$, $\lambda_2=10$, and $\lambda_3=0$). During this refinement stage, unlabeled field data are reintroduced to further adapt the model to learn and capture real-world geological complexity. To achieve computationally efficient and stable refinement, we adopt a parameter-efficient fine-tuning strategy based on Low-Rank Adaptation (LoRA) \citep{hu2021lora}, where only the LoRA parameters are updated while the remaining network's parameters are frozen. This training strategy enables lightweight refinement while preserving the foundational representations learned from previous training stages. Besides, we also add a small subset of labeled synthetic data in the training process to mitigate potential overfitting to field data and prevent catastrophic forgetting of the task-special mapping previously learned from synthetic supervision. This hybrid training process allows the network to retain its learned stratigraphic representations while progressively adapting to the complex structural patterns present in field data.

As shown in Figure~\ref{fig:fig1_3}c, the hybrid prior loss on the field dataset gradually converges to 0.154, while the loss on the small subset of synthetic data remains low at approximately 0.006 after 100 epochs of model training. This indicates that the proposed prior-driven refinement effectively improves the model’s adaptability and prediction accuracy on field data, while preserving the previously learned seismic-to-RGT mapping without noticeable degradation, thereby reducing the synthetic-to-field domain gap. As shown in the third column of Figure~\ref{fig:fig1_7}, compared to the network trained solely under synthetic supervision, the incorporation of geological prior constraints leads to significant improvements in stratigraphic continuity and structural alignment. Although the refined model generalizes well across 3,000 field datasets, it still tends to capture frequently occurring field patterns and geological structures. For individual field data with rare or highly complex geological structures, the AI model remains limited in resolving fine-scale stratigraphic and structural details.

\subsection{Domain-adaptation Fine-tuning}
\plot{fig1_6}{width=\textwidth}
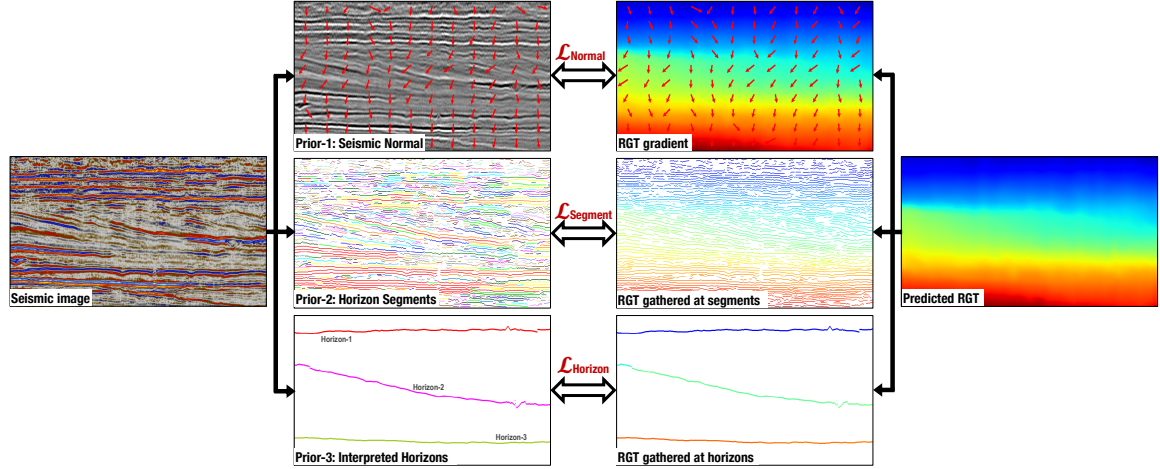
{Domain-adaptation fine-tuning using unsupervised geological prior-driven losses. We formulate automatically extracted seismic normal vectors, local horizon segments, and expert-interpreted horizons as label-independent loss functions to directly regularize RGT predictions, enforcing structural orientation alignment, local stratigraphic isochronism, and global horizon consistency.}

\plot{fig1_7}{width=\textwidth}
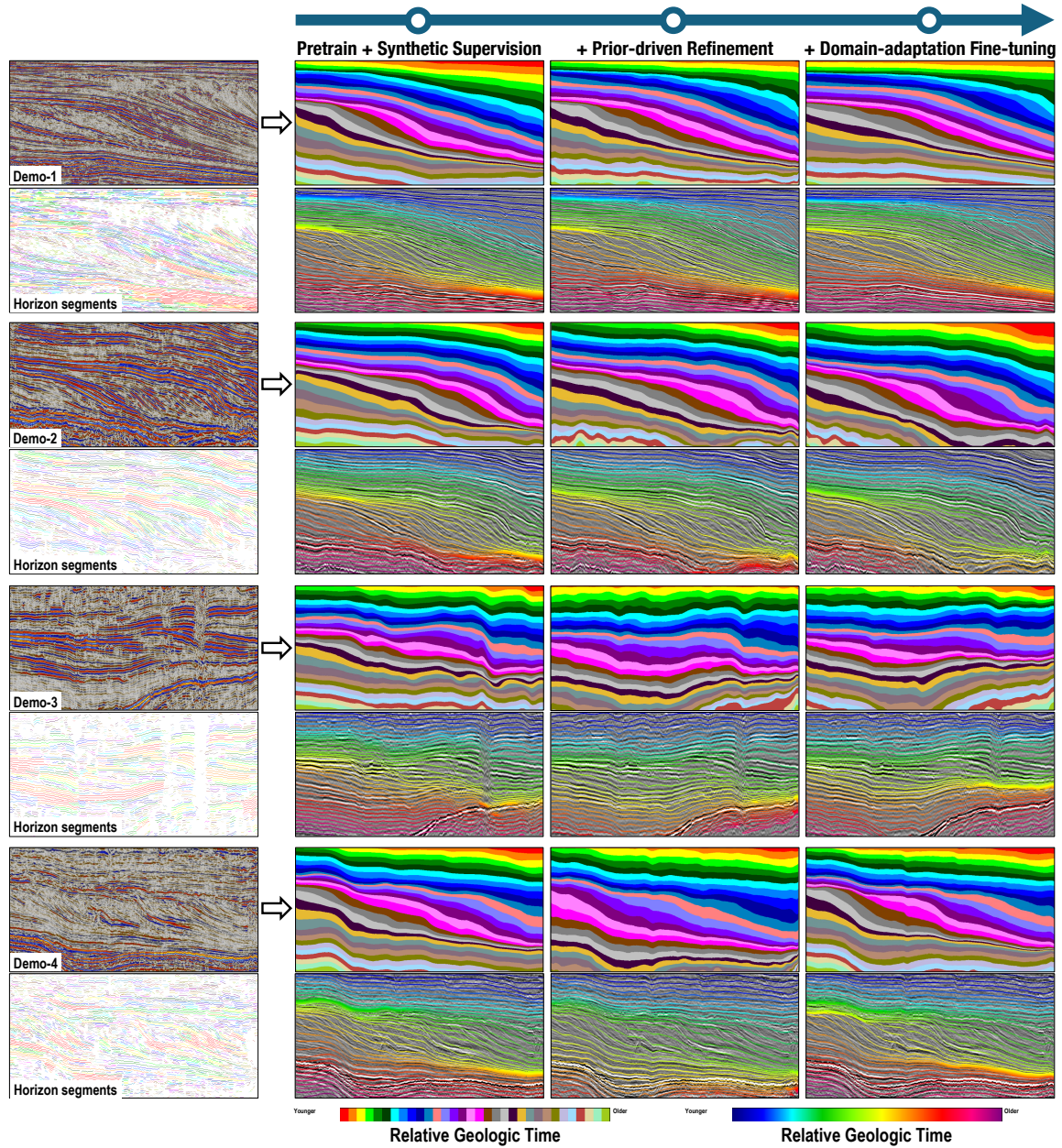
{Progressive improvement of predicted RGT results and extracted isosurfaces across the multi-stage training process, including self-supervised pretraining and synthetic supervision, geological prior-driven refinement, and sample-special effective adaptation. The initial predictions in second column capture only coarse, global structural trends and exhibit poor alignment with seismic reflectors. After incorporating geological prior-driven refinement and adaptation, the predictions in third and fourth column significantly improve the consistency with both large-scale geometries and fine-scale stratigraphic and structural variations. In contrast, the network trained without large-scale pretraining shows poor convergence and generalization, highlighting the importance of foundation model pretraining for RGT estimation in clinothems-dominated systems.}

To further improve the model‘s performance on individual field sample during practical application and deployment, we propose a domain-adaptation fine-tuning strategy that enhances the network's ability to capture and fit fine-scale stratigraphic and structural details. Instead of retraining or fine-tuning on large-scale datasets, this approach focuses on iterative optimization of a single target sample, enabling targeted refinement of both feature representation and predicted results. As shown in Figure~\ref{fig:fig1_6}, we treat the geological prior losses defined in the previous section as optimization objectives. By minimizing these unsupervised prior loss functions, the RGT prediction is progressively adjusted toward geological plausibility and structural consistency, while improving its perception and fitting of fine-scale stratigraphic geometries and local structural trends. Moreover, due to the adaptation operates on only a single sample and updates only a small number of LoRA parameters, the optimization process is computationally efficient and typically converges within a few minutes.

After applying the sample-specific adaptation, the predictions shown in the fourth column of Figure~\ref{fig:fig1_7} exhibit significant improvements in RGT estimation, including clearer stratal terminations, more distinct unconformity surfaces, improved fitting of steeply dipping strata, and enhanced alignment of fine-scale seismic reflector undulations. These improvement primarily arise from the local segment isochronous constraint, which assigns higher priority to fine-scale stratigraphic and structural variations during optimization. In this way, the network progressively evolves from capturing global coarse RGT trends to resolving detailed stratigraphic geometries that are commonly smoothed out under purely synthetic supervision. Compared with the overly smooth RGT fields (second column of Figure~\ref{fig:fig1_7}) obtained by supervised learning, the proposed prior-driven refinement combined with domain-adaptation fine-tuning preserves large-scale structural trends while recovering fine-scale stratigraphic details in complex field data.

\subsection{Applications on Global Shelf-edge Clinothems}
\plot{fig1_8}{width=\textwidth}
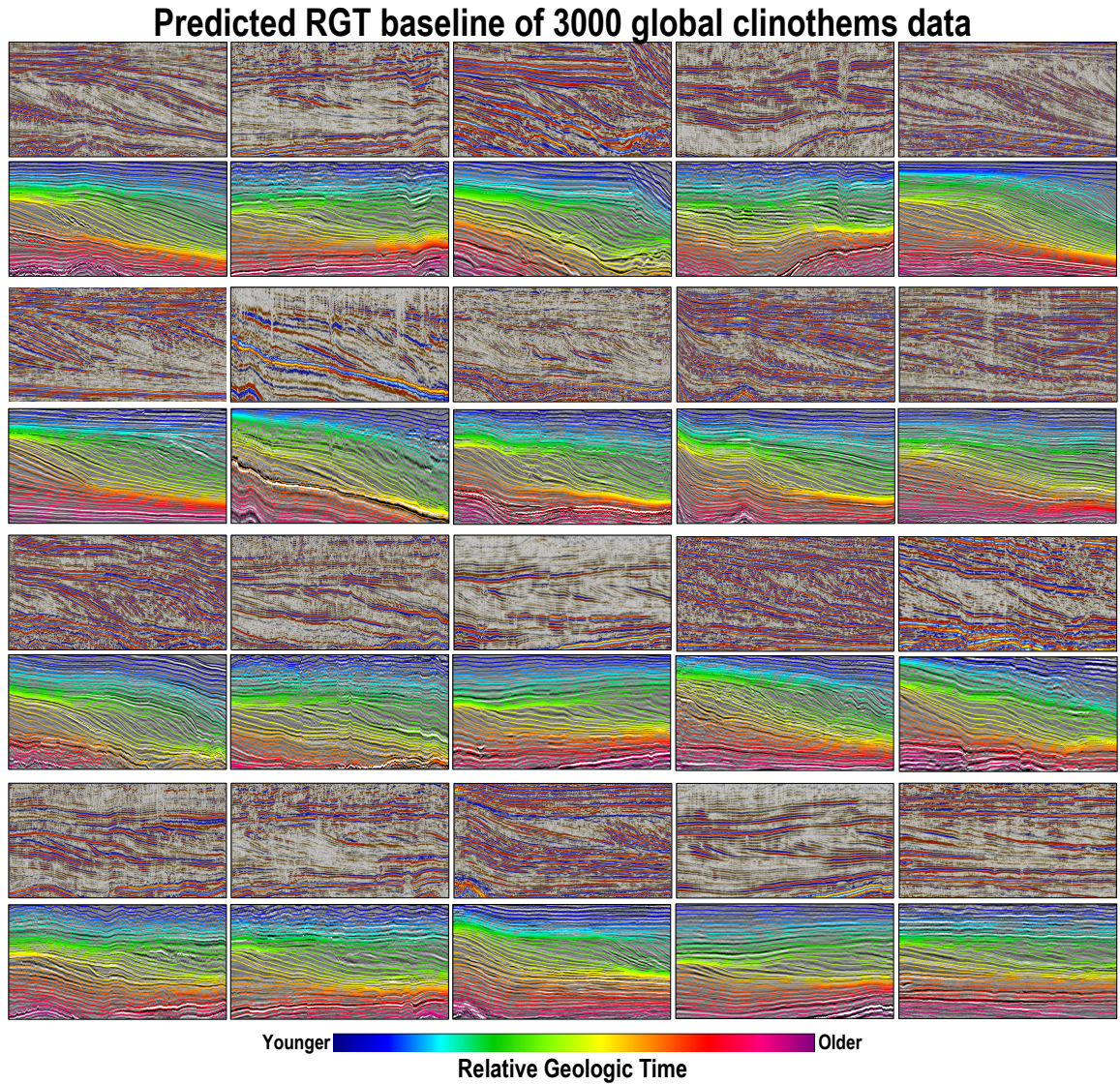
{Final RGT baseline results for global field clinothems datasets from various sedimentary basins worldwide. We apply the final model to 3,000 field datasets spanning diverse clinothems architectures, yielding accurate, robust, well-generalized, and stratigraphic and structural consistent RGT estimations.}

To evaluate the performance of the proposed geological prior-driven pretrain-to-alignment learning framework, we apply it to 3,000 field clinothems datasets (Figure~\ref{fig:fig1_8}) from multiple sedimentary basins worldwide, including the Taranaki, Northern Carnarvon, North Sea, Browse, Beagle, Offshore Canterbury, Beaufort-Mackenzie and Colville Basins. These datasets span a broad range of depositional environments, stratigraphic geometrics, and structural complexities, including various clinothems types such as sigmoidal (symmetric, divergent, and chaotic), asymmetric (top-heavy and bottom-heavy), complex, tangential (oblique and oblique chaotic), and oblique geometries \citep{anell2017quantifiable}. We randomly select several representative samples from different basins and visualize their corresponding RGT predictions in Figure~\ref{fig:fig1_8}. The predicted RGT results exhibit strong stratigraphic and structural consistency, characterized by accurate alignment with seismic reflectors, clear delineation of stratal terminations and unconformities, and robust fitting of fine-scale structural variations. Importantly, the network effectively captures fine-scale seismic reflector undulations and local structural variations, enabling pixel-level refinement that is challenging to achieve using only traditional data-driven synthetic supervision.

Finally, the proposed framework establishes a large-scale and consistent RGT baseline for 3,000 field clinothems datasets across multiple sedimentary basins. These high-resolution RGT predictions provide a dense stratigraphic framework that supports sequence stratigraphic interpretation, sedimentary evolutionary history reconstruction, reservoir property modeling, and stratigraphic and structural guidance for geophysical exploration and inversion. Moreover, these predicted RGT results for 3,000 unlabeled field datasets are publicly available as a baseline results \citep{gao2026baseline}, facilitating the development, comparison, and evolution of AI-based methods for automated seismic stratigraphic interpretation in clinothems-dominated systems.

\section{Discussion}
\plot{fig1_9}{width=\textwidth}
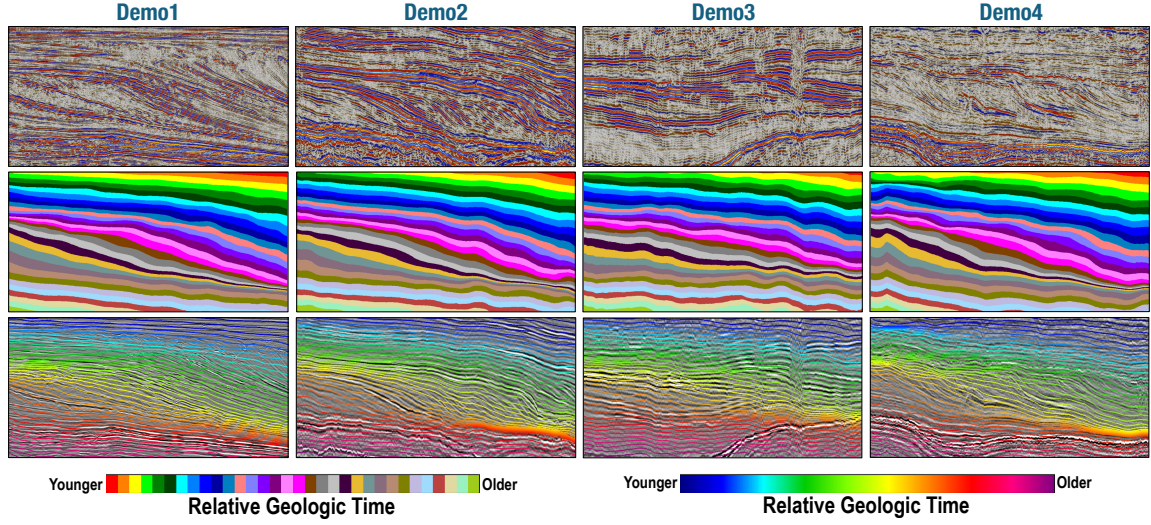
{RGT predictions obtained excluding the self-supervised pretrain stage from the pretrain-to-alignment workflow. Compared to the pretrained backbone used in the proposed framework, the non-pretrained model starts from a weaker initialization, produces overly smoothed predictions, and shows limited convergence and generalization, highlighting the importance of large-scale pretrain on seismic foundation model for RGT estimation in clinothem-dominated systems.}

Applications to the case study of RGT estimation demonstrate that the proposed geological-prior-driven pretrain-to-alignment AI framework achieves accurate, robust, and well-generalized RGT estimation across various sedimentary basins. Through the progressive integration of multiple training stages, the AI model successfully and effectively learns to capture both global structural trends and fine-scale stratigraphic details. The progressive improvements in the predictions indicate that this learning paradigm plays a critical role in guiding the network from capturing only globally smoothed RGT representations toward perceiving intrinsic stratigraphic and structural characteristics embedded in seismic data. In particular, the refinement and adaptation stages substantially improve the fitting of fine-scale stratigraphic and structural details, including stratal terminations, unconformity surfaces, and steeply dipping reflectors, which are essential for accurate clinothems interpretation in complex field datasets.

These promising results further indicate that the establishment and enhancement of model capability does not solely arise from any single training stage, but from the progressive integration of multiple stages, through which different capabilities are gradually acquired, learned, refined, and adapted. Within this framework, self-supervised pretraining provides a field-aware representational foundation that improves the model’s sensitivity to field statistical patterns and structural characteristics. Synthetic supervision then converts this representational capability into an initial task-specific mapping under plausible and controllable labels. However, these two stages remain insufficient to address complex field variability and the demands of practical deployment. Prior-driven refinement is therefore introduced to align the learned mapping with various field conditions by improving physical consistency of predictions under field complexity, ultimately reducing the synthetic-to-field domain gap. Finally, domain-adaptation fine-tuning further enhances the target-specific adaptability and deployment reliability by directly optimizing the model's predictions to adapt to sample-level complexity and variability in practical applications.

More broadly, the proposed learning paradigm has implications beyond the present RGT case study. Its methodological significance lies in providing a comprehensive and transferable learning framework for solving geophysical AI problems, in which representation learning, synthetic supervision, field alignment, and deployment adaptation are systematically integrated within a unified training workflow. This learning strategy is particularly valuable for geophysical tasks where large-scale field labels are scarce, synthetic-to-field domain gaps are pronounced, physical or geological consistency cannot be guaranteed using purely data-driven supervision, and practical deployment further requires adapting to target-specific field variability. Therefore, the proposed paradigm provides a practical methodological reference for a broader range of geophysical AI tasks, such as interpretation, regression, and inversion, under complex real-world conditions.

While the present case study confirms the effectiveness of the proposed paradigm and its broader methodological relevance, it also reveals several limitations in the current RGT-oriented implementation. Specifically, the geological priors introduced in this study, including seismic normal vectors, local horizon segments, and expert-interpreted horizons, provide only partial constraints for complex stratigraphic architectures. Future work can incorporate additional priors derived from faults, unconformities, stratigraphic boundaries, or other automatically extracted geological elements to further improve field alignment and prediction consistency. Another limitation arises from the current 2D formulation in current case study, whereas RGT estimation requires strong global consistency, making direct extension to 3D non-trivial. Extending the framework to fully 3D or computationally efficient 2.5D settings could better exploit spatial continuity and enable more global, robust, and geologically consistent stratigraphic constraints. In particular, local horizon segments extracted in 2D are inherently limited by faults and low signal-to-noise ratio regions, which restrict their lateral continuity and spatial extent, whereas 3D volumes provide richer spatial connectivity that allows horizon segments to propagate through alternative pathways, thus providing more global, robust, and geologically consistent stratigraphic constraints.

Finally, this study highlights the critical role of pretrained seismic foundation model in addressing complex regression tasks such as RGT estimation. Accurate RGT prediction of clinothems requires the robust features extraction and representation capacities that can simultaneously capture both global structural trends and fine-scale stratigraphic details.  As shown in Figure~\ref{fig:fig1_9}, excluding only the large-scale pretraining stage from the paradigm leads to predicted results that remain overly smoothed structural trends and exhibit weak stratigraphic consistency and poor fitting of fine-scale seismic reflectors. Our experiments indicate that training such a foundation model from scratch is often inefficient and unstable, whereas pretrained backbones provide a strong initialization that significantly improves both convergence behavior and generalization performance. These results further support the importance of integrating large-scale pretraining with geological prior-driven refinement and adaptation when addressing challenging seismic stratigraphic interpretation problems.

\section{Conclusion}
In this study, we propose a pretrain-to-alignment learning paradigm for improving the applicability of geophysical AI under scarce field labels and pronounced synthetic-to-field gaps. By systematically integrating self-supervised pretraining, synthetic supervision, prior-driven refinement, and domain-adaptation fine-tuning within a unified progressive workflow, the proposed paradigm provides a comprehensive and transferable learning route from field-aware representation learning to field-aligned and deployment-ready prediction. We validate this paradigm through a representative case study of RGT estimation in global shelf-edge clinothems, where the final AI model achieves accurate, robust, and well-generalized predictions across 3,000 diverse field datasets, while substantially improving the fitting of fine-scale stratigraphic and structural details. More importantly, this study highlights the methodological value of combining synthetic supervision with label-independent geological constraints, and demonstrates the broader potential of this general paradigm for geophysical interpretation, regression, and inversion scenarios. Besides, we have publicly released the baseline RGT predictions for 3,000 field datasets (https://doi.org/10.5281/zenodo.18910332,~\cite{gao2026baseline}) and corresponding source code for the proposed pretrain-to-alignment learning paradigm, facilitating the development, comparison, and evaluation of AI-based seismic stratigraphic interpretation and other related geoscience applications. 

\section{Acknowledgements}
The authors thank the USTC supercomputing center for providing computational resources, and CIGVis \cite{li2025cigvis} for providing data visualization approach for this project.


\section{Conflict of Interest}
The authors declare no conflict of interest.

\section{Data and Materials Availability}
The baseline RGT predictions for 3,000 field datasets have been uploaded to Zenodo and are freely available at https://doi.org/10.5281/zenodo.18910332 \citep{gao2026baseline}. The corresponding source codes in this study have been uploaded to GitHub, and are freely available for further research.

\bibliographystyle{plainnat} 
\bibliography{huig26global}

\clearpage
\end{document}